\documentclass{Interspeech2024}

\interspeechcameraready

\title{1000 African Voices: Advancing inclusive multi-speaker multi-accent speech synthesis}

\name[affiliation={1,*}]{Sewade}{Ogun}
\name[affiliation={2,*}]{Abraham T.}{Owodunni} 
\name[affiliation={2,*}]{Tobi}{Olatunji}
\name[affiliation={3,*}]{Eniola}{Alese}
\name[affiliation={3,*}]{Babatunde}{Oladimeji}
\name[affiliation={4,5,*}]{Tejumade}{Afonja}
\name[affiliation={6,*}]{Kayode}{Olaleye}
\name[affiliation={7,*}]{Naome A.}{Etori}
\name[affiliation={8,*}]{Tosin}{Adewumi}

\address{\small
  $^1$Université de Lorraine, CNRS, Inria, LORIA, F-54000 Nancy, France
  $^2$Intron Health
  $^3$Amazethu Research
  $^4$CISPA Helmholtz Center for Information Security
  $^5$AI Saturdays Lagos
  $^6$Data Science for Social Impact Group, University of Pretoria
  $^7$University of Minnesota - Twin Cities, USA
  $^8$Luleå University of Technology
  $^*$Masakhane NLP}
\email{\small sewade.ogun@inria.fr, tobi@intron.io}


\begin{document}

\maketitle


\begin{abstract}
    
Recent advances in speech synthesis have enabled many useful applications like audio directions in Google Maps, screen readers, and automated content generation on platforms like TikTok. However, these systems are mostly dominated by voices sourced from data-rich geographies with personas representative of their source data. Although 3000 of the world's languages are domiciled in Africa, African voices and personas are under-represented in these systems. As speech synthesis becomes increasingly democratized, it is desirable to increase the representation of African English accents. We present Afro-TTS, the first pan-African accented English speech synthesis system able to generate speech in 86 African accents,
with 1000 personas representing the rich phonological diversity across the continent for downstream application in Education, Public Health, and Automated Content Creation. Speaker interpolation retains naturalness and accentedness, enabling the creation of new voices. 
\end{abstract}

\section{Introduction}

Synthetic voices are used in everyday applications for text reading, content generation, voice-over, etc., and provide audio feedback in applications such as maps, language learning tools, smart speakers, and voice assistants. Synthetic voices have also found wider adoption as a plugin with the proliferation of large language models. With the high naturalness and high quality of synthetic voices \cite{kim2021conditional, wang2023neural}, they have become more widely used on online platforms and social media. 
 
Synthetic voices are usually generated by speech synthesis systems, and there are several open-source systems available for use, e.g., \cite{casanova2022yourtts}. Several of these systems are provided in English, and several efforts have been made to cover over 1000 languages of the world, including several African languages \cite{pratap2023scaling}.

Over 1.4 billion people reside in Africa, and more than 237 million people speak the English language at a bilingual or native level \cite{enwiki:1211690147}, including Nigeria, Uganda, South Africa, etc., and several other African countries speak it as a second or third language. Considering this large demography of speakers, the current representation of personas in synthetic voices does not show this diversity in terms of the typical African accent. 

Generating synthetic speech with personas similar to one's demography is paramount for the widespread adoption and acceptability of the technology. For example, a Nigerian content creator would prefer to generate speech in an accent (or language) that is familiar and natural to their audience. Voice cloning methods have been created to generate speech in the voice of a reference speaker, however, these systems still fail to generalize to the diverse African accents from our analysis. This is because speaker representation in speech synthesis systems favors more general accents like the American, British, or Australian accents, etc. Even the widely used English text-to-speech (TTS) datasets are not representative of diverse demographies \cite{ljspeech17, yamagishi2019cstr, zen2019libritts}. 

Therefore, in this work, we focus on expanding the diversity of synthetic personas in TTS systems to the typical African accent. Our work covers 747 speakers and 86 accents from 9 countries. We focus on improving several TTS systems in generating voices with African-like personas, enabling a larger representation of African voices for use in applications like podcasts and content creation.
Finally, we explore a speaker averaging method to create new personas different from those used to train the TTS systems. This was done to ensure that users can generate synthetic speech of diverse accents, without the fallout of overuse of a specific speaker or accent in our dataset.

Section~\ref{review} reviews some literature particular to this research. Sections~\ref{methodology} and \ref{experiment} describe the dataset, data preprocessing methods, TTS models, evaluation protocols, and experiments. We discuss the results in Section~\ref{results} and conclude with limitations and the summary of our work in Sections~\ref{limitations} and \ref{conclusion} respectively.

\section{Literature Review}
\label{review}

There have been works extending TTS to other accents. These include works on multilingual, multi-speaker TTS synthesis, e.g., \cite{casanova2022yourtts}, where the accents of different speakers were learned jointly with the language, enabling the model to produce English speech of different accents. Similarly, \cite{badlani2023multilingual} disentangled the speaker from the accent so that the accent can be assigned to any speaker in the TTS model. Some curated datasets also covered African languages, e.g., CMU Wilderness \cite{black2019cmu} and MMS \cite{pratap2023scaling} datasets, but majorly have single speakers for each language.

In the African context, the authors of \cite{ogayo22_interspeech} curated datasets and built speech synthesis systems for 12 low-resourced African languages. Similarly, BibleTTS \cite{meyer22c_interspeech} contained 86 hours of recordings involving 10 African languages, including Asante Twi, Hausa, and Yoruba. The authors in \cite{49562} introduced an open-source corpus of Yoruba, a language spoken by over 22 million people. Other efforts include Lagos-NWU Yoruba corpus \cite{van2015lagos} involving 16 female and 17 male speakers, and the work by \cite{van2012tone} involving 33 speakers. Gamayun \cite{oktem2020gamayun} introduced multi-speaker TTS for some marginalized languages and code-switched data for four South African languages by \cite{van2017rapid}. Similarly, a TTS model was implemented in Festival for the Fon language by \cite{dagba2014text}. Although these works have explored creating TTS datasets and TTS systems for African languages, our work focuses on increasing the representation of African accents in English-based 
TTS systems. 

\section{Methodology}
\label{methodology}
\subsection{Dataset: Afro-TTS}
\label{dataset}

Curating a dataset of diverse African-accented English speakers is crucial for this work. Therefore, we initiated our research by online crowd-sourcing of data, following the methodology used in a previous work \cite{olatunji2023afrispeech}. 
Volunteers were asked to read and record a sequence of English texts. 
The texts contained general domain words and were specifically enriched with African-named entities to accurately represent the diversity of African names and organizations. Table~\ref{table:data_stats} shows the dataset statistics. The data collection process involved 747 paid contributors from 9 countries representing 86 accents. Contributors consented to have their audio used for TTS. The 136-hour, 16-bit, 48~kHz audio files, along with their metadata containing anonymized speaker identity, country, accent, age group, gender, etc., have been open-sourced to facilitate African speech research.\footnote{models and dataset can be accessed via \url{https://huggingface.co/intronhealth/afro-tts}}

\begin{table}[t]
\centering \caption{Afro-TTS dataset statistics. The country column indicates the ISO~3166-1 country code.}
\label{table:data_stats}
\addtolength{\tabcolsep}{-2pt}
\begin{tabular}{l c c c c}  
\hline
Country & \# samples & \# speakers & \# accents & Duration (h) \\ \hline
NG & 25564 & 549 & 48 & 99.85 \\
KE & 5307 & 58 & 8 & 16.45 \\
ZA & 4279 & 125 & 20 & 16.41 \\
GH & 727 & 4 & 3 & 2.28 \\
ZW & 47 & 6 & 3 & 0.20 \\
RW & 40 & 1 & 1 & 0.15 \\
SL & 38 & 1 & 1 & 0.14 \\
UG & 26 & 2 & 1 & 0.08 \\
ZM & 8 & 1 & 1 & 0.04 \\
\hline
\vspace{-20pt}
\end{tabular}
\end{table}
\subsubsection{Dataset preprocesing}

As the dataset was recorded remotely on different devices, it needed to be processed to be suitable for TTS training. Firstly, the speech samples were denoised using a speech enhancement model \cite{defossez20_interspeech}, which removes various background noise, including stationary and non-stationary noises, and room reverberation. The denoised samples were then passed through a bandwidth-extension model, VoiceFixer \cite{liu2021voicefixer}, to improve the quality of some of the highly degraded utterances, i.e., utterances recorded with low-resolution and clipping distortion. VoiceFixer has three audio processing modes which are suitable for different levels of degradation. To keep the audio file with the best quality per sample, we evaluated the quality of the denoised sample and the three enhanced samples using a quality estimator, WVMOS \cite{hifi++, ogun2023can}\footnote{\url{https://github.com/AndreevP/wvmos}}, then we selected the sample with the highest predicted MOS score among the samples per utterance. All samples were processed at a sampling rate of 16~kHz.
The speech samples were then normalized to a volume level of -27~dB using the RMS-based normalization method in ffmpeg\footnote{\url{https://github.com/slhck/ffmpeg-normalize}} and a voice activity detection tool \cite{py-webrtcvad} was used to eliminate long pauses from the speech recordings. 

We selected 736 samples covering speakers and accents with more than 20 minutes of data for testing. Finally, the remaining samples were randomly split into training (35042) and development (200) sets. We excluded recordings with a duration of over 50 seconds or samples with character counts exceeding 400 for faster training.

Abbreviations and name titles were expanded, e.g., ``Alh~$\rightarrow$~Alhaji'', ``Maj $\rightarrow$ Major'' and
numbers were converted to their word forms using the NeMo text normalization toolkit \cite{zhang21ja_interspeech}. In addition, some punctuation marks (open
bracket, closed bracket, colon, and semicolon), were also expanded to their word form, as they were read out in the dataset according to annotation instructions.

\subsection{TTS models}

Two state-of-the-art, open-source, end-to-end TTS models, \emph{VITS} \cite{kim2021conditional} and \emph{XTTS} \cite{xttsnote}, were used in our experiments. VITS is an end-to-end model that adopts variational inference augmented with normalizing flows and an adversarial training process. 
XTTS is a recent multilingual TTS system with cross-language voice cloning capabilities comprising three modules; a VQ-VAE module, a GPT module, and an audio decoder. 
VITS (86.6~M parameters) was trained on the VCTK dataset (44~h of 109 native English speakers) for 500k iterations while XTTS (version~2, 750~M parameters) had been trained on a dataset of over 16k hours comprising 16 languages,\footnote{\url{https://docs.coqui.ai/en/latest/models/xtts.html}} including English.

The models were fine-tuned on the training set as \emph{VITS-FT} and \emph{XTTS-FT}\footnote{Only the GPT module was fine-tuned in our experiments.} respectively. We also trained a randomly initialized VITS model from scratch as \emph{VITS-O} on the training set to validate the quality of the data for TTS experiments. Lastly, we modified the VITS model as \emph{VITS-EXT} to take an external 256-dimensional l2-normalized speaker embedding vector as speaker conditioning. The speaker embeddings were extracted from Resemblyzer \cite{resemblyzer}, a speaker embedding extractor. The weights of incompatible layers were re-initialized during fine-tuning, e.g., the speaker embedding module.

\begin{table*}
\centering
\caption{Objective and subjective evaluation results (with 95~\% confidence interval) for pre-trained and fine-tuned TTS models. Mean opinion score (MOS), naturalness MOS (Nat-MOS), model-based MOS (WV-MOS), model-based naturalness (NISQA), cosine similarity (cos-sim), accentedness MOS (Accent-MOS), \%~word error rate (\%~WER), and preference rankings are provided.} 
\label{table:model_results}
\addtolength{\tabcolsep}{-2.7pt}
\begin{tabular}{ |l|  c  c|  c  c|  c  c | c | c |}
\cline{2-9}
 \multicolumn{1}{l|}{}& \multicolumn{4}{c|}{Overall quality and naturalness} & \multicolumn{2}{c|}{Speaker \& accent similarity} & Intelligibility & Preference \\
\hline
 Model & MOS & Nat-MOS & WV-MOS & NISQA & cos-sim & Accent-MOS & \% WER & Ranking \\
\hline
GT denoised &  $3.75\pm0.04$ & $4.55\pm0.03$ & $2.85\pm0.04$  & $4.55\pm0.03$ & - & $4.49\pm0.03$ & - & - \\
\hline
VITS & $3.80\pm0.04$ & $2.84\pm0.06$ & $4.26\pm0.02$ & $3.84\pm0.04$  & - & $1.81\pm0.05$ & $32.75$ & - \\
XTTS & $3.92\pm0.04$ & $3.31\pm0.06$ & $3.71\pm0.02$ & $3.00\pm0.03$  & $0.828$ & $2.31\pm0.06$ & $13.81$ & - \\
\hline
\multicolumn{9}{c}{Trained/fine-tuned models} \\
\hline
VITS-O & $3.02\pm0.05$ & $4.00\pm0.04$ & $2.93\pm0.03$ & $2.94\pm0.02$ & $0.834$ & $4.02\pm0.04$ & $66.77$ & - \\
VITS-FT & $3.33\pm0.04$ & $4.18\pm0.04$ & $3.03\pm0.03$ & $2.97\pm0.03$ & $0.834$ & $4.16\pm0.04$ & $51.77$ & (1192) 2rd \\
VITS-EXT & $3.14\pm0.05$ & $4.07\pm0.04$ & $3.02\pm0.03$ & $\mathbf{3.06\pm0.03}$ & $\mathbf{0.914}$ & $4.07\pm0.04$ & $57.31$ & (1168) 3rd \\
XTTS-FT & $\mathbf{3.77\pm0.04}$ & $\mathbf{4.39\pm0.03}$ & $\mathbf{3.31\pm0.03}$ & $\mathbf{3.07\pm0.04}$ & $\mathbf{0.889}$ & $\mathbf{4.35\pm0.03}$ & $\mathbf{19.20}$ & \textbf{(1235) 1st} \\
\hline
\end{tabular}
\vspace{-15pt}
\end{table*}

\subsection{Speaker interpolation}

Speaker interpolation has been typically used in Hidden Markov Model (HMM)-based TTS systems for changing speaker characteristics \cite{yoshimura1997speaker}. Interpolation can be done in several ways including interpolating between different Gaussian distributions of speakers or between speaker representations. Given speakers $S1$ and $S2$, for example, a new speaker $S3$ can be generated using a linear interpolation of their speaker representations:
\begin{equation}
 S3 = \alpha*S1 + (1-\alpha)*S2
\end{equation}
The interpolation ratio $\alpha$ enables the speaker characteristics to be changed from one speaker to another along a spectrum. Therefore, to generate more personas with African accents, we interpolated speakers with the same gender, country, and accent, using their speaker embeddings, creating over 200 additional speakers. We filtered by country because there are regional differences in the accent from different countries even for the same language, e.g., Swahili and Hausa language speakers in different countries. 

\subsection{Evaluation protocol}

Several subjective and objective metrics were used to compare the performance of the TTS systems. For objective evaluation, we computed the overall quality \textbf{(WV-MOS)} and the naturalness \textbf{(NISQA)} \cite{mittag2020deep} of the utterances using model-based quality estimators. 
The cosine similarity \textbf{(cos-sim)} of the target speaker embedding to the reference speaker embedding was also computed to measure the speaker similarity. To measure the intelligibility of the generated utterances, we computed the word error rate \textbf{(WER)} between the ground truth text and text predicted by Whisper-large-v3 \cite{radford2023robust} from the Hugging Face library.

For subjective evaluation, we performed listening tests on the generated utterances using the Intron Speech Vault Platform.\footnote{\url{https://speech.intron.health}} We asked participants to evaluate the overall quality \textbf{(MOS)}, naturalness \textbf{(Nat-MOS)}, and accentedness \textbf{(Accent-MOS)} of the utterances on a Likert scale of 1--5 with 1.0 intervals. MOS evaluates the lack of noise and correct pronunciation while Nat-MOS measures how natural or human-like the speech sounds, regardless of intelligibility or clarity.
We also asked participants from the same country as the reference accent to rate the closeness of the generated utterance to the reference accent \textbf{(Accent-Match)}, reference country \textbf{(Country-Match)}, and reference gender \textbf{(Gender-Match)} on a scale of 1--5. 
Finally, we performed a \textbf{preference} test\footnote{\url{https://huggingface.co/spaces/eniolaa/AfroTTS-Evaluation}} on our accented models using a modified Hugging Face TTS Arena tool \cite{ttsarena}. Here, volunteers were shown a pair of utterances generated by two TTS models among the finetuned TTS models at every turn, then they were asked to select the more natural utterance between the pair.

In addition, the naturalness of utterances generated by the newly created speakers was evaluated objectively and through listening tests. We also computed the equal error rate \textbf{(EER)} of the new speaker's utterance against the utterances of the averaged speakers to validate that they are indeed different. 

\section{Experiments}
\label{experiment}

\subsection{Training and finetuning hyper-parameters}

All the VITS-like models, VITS-O, VITS-FT, VITS-EXT, were trained with mixed-precision training on 4~GPUs for 350k iterations, with a global batch size of 64. We used the VITS implementation from the authors\footnote{\url{https://github.com/jaywalnut310/vits}} for training and inference, with the default training hyper-parameters. Similarly, XTTS was fine-tuned on 1~GPU using the Coqui library\footnote{\url{https://github.com/coqui-ai/TTS}} for about 250 iterations with default finetuning hyper-parameters, using a batch size of 2 and gradient accumulation of 128. 
The VITS-like models take phonemes (without stress markers) interspersed with a blank token as input and generate audio at a 16~kHz sampling rate while the XTTS-FT model was fine-tuned at 24~kHz by upsampling the 16~kHz processed training data. In addition, to mitigate regional bias during model training, we computed the average duration per speaker on the original dataset, and then we duplicated all the samples corresponding to a speaker to equal the average duration of 10 minutes if the speaker's duration was less than the average.

At inference time, we set the noise scale of the VITS-like models to 0.667, with the duration noise scale set to 0.8 following the original work. For speaker and accent representation, we either used the learned speaker embedding, the speaker embedding extracted from the speaker extractor model (for VITS-E), or the test utterance (for XTTS models). XTTS-based models generate speech at 24~kHz while VITS-like models generate speech at 16~kHz. Therefore, all the generated samples were resampled to 16~kHz for evaluation.
Also, the speaker interpolation parameter $\alpha$ was set to 0.5, and at most three speakers were interpolated.
Additionally, statistical significance tests were carried out on the presented results using the bootstrap method. The best model or best models with similar scores among the trained and fine-tuned models are highlighted in bold.

\begin{table}
    \centering
        \caption{Country-level MOS results (with 95~\% confidence interval) for the best model (XTTS-FT) showing naturalness (Nat-MOS), accentedness (A-MOS), country-match (Country-M), and accent-match (Accent-M).}
    \addtolength{\tabcolsep}{-2.5pt}
    \label{tab:xtts_ft_mos_by_country}
    \begin{tabular}{lccccc}
\hline
{} & Nat-MOS & A-MOS &  Country-M &   Accent-M  \\
\hline
KE         &  $4.44\pm0.09$ &        $4.37\pm0.09$ &  $3.93\pm0.52$ &   $3.90\pm0.48$\\
NG      &  $4.37\pm0.04$ &        $4.33\pm0.04$ &  $4.24\pm0.11$ &  $3.54\pm0.15$\\
ZA  &  $4.46\pm0.11$ &         $4.47\pm0.10$ &   $3.40\pm0.49$ &  $2.93\pm0.53$\\
\hline
\end{tabular}
\vspace{-15pt}
\end{table}

\section{Results and Discussion}
\label{results}

Aggregated results for objective and subjective evaluations are presented in Table~\ref{table:model_results} excluding results for speaker interpolated speakers. Furthermore, in Tables~\ref{tab:xtts_ft_mos_by_country} and~\ref{tab:xtts-accent-level}, we show a breakdown of the ratings for the best fine-tuned model where the raters' country/accent match the utterance country/accent enabling us to measure the similarity in accent, country, and gender of the utterances. Here, we only show results for three counties where we had significant evaluators to measure the similarity in accent, country, and gender of the utterances. Also, these three countries make up 97.8~\% of the dataset.

We generated 735 utterances per model, along with 162 speaker-interpolated utterances from VITS-based models.
Also, 427 unique participants (64.6~\% female, 60 accents, from 9 countries) provided 26,974 human ratings representing roughly 5 ratings per utterance. Additionally, 26,116 ratings were provided across models and 858 ratings were provided for speaker-interpolated utterances. \emph{GT denoised} are the Ground truth (GT) reference test samples. 

\subsection{Naturalness and overall quality results}
Table~\ref{table:model_results} shows that although participants rated the pre-trained models higher in overall quality (MOS), the best fine-tuned model (XTTS-FT) was rated 1.08 MOS points higher than its pre-trained version, and 1.55 MOS points higher than the VITS baseline in terms of naturalness, probably due to the better pronunciation of African named entities in the reference text.
Our results demonstrate that we are indeed able to generate speech that is more relatable to an African audience. However, model-based quality metrics (WV-MOS and NISQA) show inverse results (pre-trained models are better) possibly because underlying models lack exposure to African-sounding speech.

\begin{table}
    \centering
        \caption{Accent-level: Best model (XTTS-FT) results for ratings where the utterance accent is matched to the rater's accent.}
    \label{tab:xtts-accent-level}
    \begin{tabular}{ccc}
\hline
Accent &  Country-Match &   Accent-Match \\
\hline
Afrikaans             &   $4.80\pm0.56$ &  $2.67\pm5.17$ \\
Hausa                 &  $4.23\pm0.22$ &  $3.93\pm0.25$ \\
Igbo                  &  $4.12\pm1.37$ &  $2.25\pm1.24$ \\
Swahili               &  $3.89\pm0.56$ &  $3.77\pm0.57$ \\
Tswana                &  $3.75\pm3.01$ &   $3.50\pm1.59$ \\
Yoruba                &  $4.25\pm0.14$ &    $3.30\pm0.20$ \\
Zulu                  &  $3.09\pm0.59$ &  $2.88\pm0.62$ \\
\hline

\end{tabular}
\vspace{-15pt}
\end{table}

\subsection{Accentedness and speaker similarity results}
Most significantly in Table~\ref{table:model_results}, participants rated XTTS-FT only 0.14 MOS points lower in Accentedness (Accent-MOS) than the GT in contrast to the XTTS baseline which was rated 2.18 MOS points below the GT. This validates that our approach generates natural-sounding accented speech, bridging the current gap in the representation of African voices in speech synthesis. Speaker similarity results (cos-sim) also showed that generated utterances from fine-tuned models are closer to reference utterances than generated utterances from pre-trained models.

\subsection{Preference scores}
Preference test scores in Table~\ref{table:model_results} show that raters prefer utterances generated by XTTS-FT.
Preference test scores aligned well with MOS metrics where the XTTS-FT model outperformed the VITS-based models. Also, the pre-trained XTTS model had a higher MOS score on average than the pre-trained VITS model likely because of its multilingual pretraining.

\subsection{Intelligibility}
Utterances by XTTS models had lower WER than VITS-based models perhaps as a result of the greater diversity and quantity (363x) of its pretraining data. A higher WER after fine-tuning of XTTS was also due to noise artifacts in the Afro-TTS dataset.

\subsection{Regional diversity considerations}
Table~\ref{tab:xtts_ft_mos_by_country} reveals that although the naturalness and accentedness of generated utterances from our best model are close to the GT, regional differences surface. South Africans (ZA) rated South African-generated utterances lower in Accent-Match than West Africans (NG) rated the generated utterances with West African accents. Although most participants agreed that the generated utterances represent the reference country from Table~\ref{tab:xtts-accent-level}, the generated accents do not always match the reference accent, e.g., generated speech in Afrikaans accent may sound like Zulu, and Igbo generated accent may sound like Yoruba. Indeed, in multilingual countries like NG and ZA, speaker accents are difficult to classify into binary accent classes \cite{owodunni-etal-2024-accentfold}, as many speakers have dual accents.
Notably, although East African speakers have lower representation in the dataset compared to Southern Africans, East Africans (e.g., Swahili) generally rated Accent-Match higher than South Africans (e.g., Zulu, Afrikaans). These inconsistencies may reflect the accent imbalance in the Afro-TTS dataset and require further investigation. 

\begin{table}
    \centering
        \caption{Speaker interpolation results showing MOS, naturalness (Nat-MOS), and accentedness (A-MOS) of utterances generated using novel speakers from speaker interpolation.}
    \label{table:results_spk_interpolation}
    \addtolength{\tabcolsep}{-3.0pt}
    \begin{tabular}{lcccc}
\hline
Model & MOS &    Nat-MOS & A-MOS & \%EER \\
\hline
VITS-EXT &   $3.17\pm0.16$ &  $4.05\pm0.15$ &  $4.10\pm0.14$ & $20.42$ \\
VITS-FT  &   $3.47\pm0.15$ &  $4.22\pm0.14$ &  $4.37\pm0.11$ & $16.20$ \\
VITS-O   &   $3.18\pm0.17$ &  $4.08\pm0.16$ & $4.20\pm0.13$ & $16.20$ \\
\hline
\end{tabular}
\vspace{-5pt}
\end{table}

\subsection{Effects of speaker interpolation}
Table~\ref{table:results_spk_interpolation} shows MOS results on speaker-interpolated utterances from fine-tuned VITS models. Although \%EER shows interpolated speakers have a high correlation with source speakers, our results show that speaker interpolation is indeed a viable approach for creating novel synthetic speakers that sound African (i.e., natural and accented). Furthermore, Table~\ref{tab:results_spk_interpolation_gender} shows that the generated utterances' gender, accent, and country match that of the reference interpolated speakers. In the future, speaker interpolation outside of the same accents could facilitate the exploration of novel or multilingual accents. 

\begin{table}
    \centering
    \small        \caption{Overall speaker interpolation MOS results showing how much synthetic utterances from interpolated speakers match the expected accent, country, and gender of the source speakers. Accent-match (Accent-M), gender-match (Gender-M), and country-match (Country-M) are provided.}
    \label{tab:results_spk_interpolation_gender}
    \begin{tabular}{lccc}
\hline
Country &  Accent-M &   Gender-M &  Country-M  \\
\hline
KE         &  $4.17\pm0.21$ &  $4.79\pm0.13$ &   $4.20\pm0.21$ \\
NG       &  $3.65\pm0.13$ &  $4.62\pm0.08$ &  $3.91\pm0.13$ \\
ZA  &  $3.47\pm0.19$ &  $4.65\pm0.09$ &  $3.64\pm0.18$ \\
\hline
\end{tabular}
\vspace{-15pt}
\end{table}

\section{Limitations}
\label{limitations}
Although we included 86 distinct African accents, this is a small fraction of more than 3000 languages and accents across the continent. Additionally, imbalanced accent representation in our dataset may yield biased performance favoring majority accents. Lastly,  we acknowledge the privacy risk of releasing multi-speaker TTS systems that mimic voices in the source data increasing the risk of voice cloning or voice theft. We mitigate this by removing any speaker identifiers, making it more challenging to identify individuals. Finally, disentangling of speaker and accent characteristics is left for future work. 

\section{Conclusion}
\label{conclusion}
We developed an African-accented TTS system that achieves near-GT MOS for naturalness and accentedness using the pan-African TTS dataset, a 136-hour dataset containing 747 speakers with 86 African accents from 9 countries. Although open questions remain, our work greatly improves the representation of African voices in speech synthesis.

\section{Acknowledgements}

We appreciate the over 700 African contributors whose voices made this work possible. We appreciate the invaluable support from Intron Health for contributing the datasets for this work, the pan-African platform for data collection, developing custom UIs for human evaluation (MOS), quality reviews, multi-currency contributor payments, and compute for experiments. Experiments presented in this paper were partly carried out using the Grid’5000 testbed, supported by a scientific interest group hosted by Inria and including CNRS, RENATER and several Universities as well as other organizations (see \url{https:// www.grid5000.fr}). Tejumade Afonja is partially supported by ELSA – European Lighthouse on Secure and Safe AI funded by the European Union under grant agreement No. 101070617. We appreciate the support provided by the BioRAMP researchers, whose collaboration and insights have been fundamental to our research.

\bibliographystyle{IEEEtran}
\bibliography{main}

\end{document}